\documentclass[preprint,preprintnumbers,amsmath,amssymb,superscriptaddress,11pt,graphicx]{revtex4}

\usepackage{graphicx}
\usepackage{dcolumn}
\usepackage{bm}
\usepackage{amssymb}

\newcommand{\be}{\begin{eqnarray}}
\newcommand{\ee}{\end{eqnarray}}
\newcommand{\ba}{\begin{eqnarray}}
\newcommand{\ea}{\end{eqnarray}}

\renewcommand{\L}{{\cal L}}

\begin{document}

\title{Quintessence models with an oscillating equation of state and their potentials}

\author{Wen Zhao}
\affiliation {Department of Physics, Zhejiang University of Technology, Hangzhou, 310014, People's Republic of China}



\begin{abstract}

In this paper, we investigate the quintessence models with an
oscillating equation of state (EoS) and their potentials. From the
constructed potentials, which have the EoS of
$\omega_{\phi}=\omega_0+\omega_1\sin z$, we find they are all the
oscillating functions of the field $\phi$, and the oscillating
amplitudes are decreasing (or increasing) with $\phi$. From the
evolutive equation of the field $\phi$, we find this is caused by
the expansion of the universe. This also makes that it is very
difficult to build a model whose EoS oscillates forever. However
one can build a model with EoS oscillating for a period. Then we
discuss three quintessence models, which are the combinations of
the invert power law functions and the oscillating functions of
the field $\phi$. We find they all follow the oscillating EoS.

\end{abstract}



\maketitle


\section{Introduction}

Recent observations on the Type Ia Supernova (SNIa)\cite{sn},
Cosmic Microwave Background Radiation (CMB)\cite{map} and Large
Scale Structure (LSS)\cite{sdss} all suggest that the universe
mainly consists of dark energy (73\%), dark matter (23\%) and
baryon matter (4\%). How to understand the physics of the dark
energy is an important issue, which has the EoS of $\omega<-1/3$,
and leads to the recent accelerating expansion of the universe.
Several scenarios have been put forward as a possible explanation
of it. A positive cosmological constant is the simplest candidate,
however it needs the extreme fine tuning to account for the
observations. As the alternative to the cosmological constant, a
number of dynamic models have been
proposed\cite{phantom,kessence,quintom,yangmills}. Among them, the
quintessence is the most natural model\cite{quint}, in which the
dark energy is described by a scalar field $\phi$ with lagrangian
density $\L_{\phi}=\frac{1}{2}\dot{\phi}^2-V(\phi)$. These models
can naturally give the EoS with $-1\leq\omega_{\phi}\leq1$.
Usually, people discuss these models with monotonic potential
functions i.e. the models with the exponential potentials and
invert power law potentials. These models have some interesting
characters, such as some models have late time attractor solutions
with $\omega_{\phi}<0$\cite{simply}, and some have the track
solutions, which can naturally answer the cosmic ``coincidence
problem"\cite{track_quint}. Recently, a number of authors have
considered the dark energy with oscillating EoS in the
quintessence models\cite{scott}, in the quintom models\cite{osci},
in the ideal-liquid models and in the scalar-tensor dark energy
models\cite{s-t}. They discussed that this kind of dark energy may
give a naturally answer for the ``coincidence problem" and
``fine-tuning problem". And in some models, it is a naturally way
to relate the very early inflation and recent accelerating
expansion. The most interesting is that these models are likely to
be marginally suggested by some observations\cite{observation}.

In this paper, we will mainly discuss the quintessence models with
the oscillating EoS. First, we construct the potentials from the
parametrization $\omega_{\phi}=\omega_0+\omega_1\sin z$. We find
these potentials are all the oscillating functions, and the
oscillating amplitudes are increasing (decreasing) with the field
$\phi$. This character can be analyzed from the evolutive equation
of $\phi$. This suggests the way to build the potential functions
which can follow the oscillating EoS. Then we discuss three kinds
of potentials, which are the combinations of the invert power law
functions and the oscillating functions, and find that they indeed
give the oscillating EoS.

The plan of this paper is as follows: in Section 2, using the
parametrized EoS $\omega_{\phi}=\omega_0+\omega_1\sin z$, we build
their potentials, and investigate their general characters by
discussing the kinetic equation of the quintessence field; then we
build three kinds of models, and discuss the evolutions of their
potentials, EoS and energy densities in Section 3; at last we will
give a conclusion in Section 4.

We use the units $\hbar=c=1$ and adopt the metric convention as
$(+,-,-,-)$ throughout this paper.

\section{Construction of the potentials}

First, we will study the general characters of the potentials, which
can follow the oscillating EoS. We note that many periodic or
nonmonotonic potentials have been put forward for dark energy, but
rarely give rise to the periodic $\omega_{\phi}(z)$. As one
well-studied example, the potential for a pseudo-Nambu Goldstone
boson (PNGB) field\cite{pngb} can be written as
$V(\phi)=V_0[\cos(\phi/f)+1]$, clearly periodic, where $f$ is a
(axion) symmetry energy scale. However, unless the field has already
rolls through the minimum, the relation $\omega_{\phi}(z)$ is
monotonic and indeed can well described by the usual
$\omega_{\phi}(a)=\omega_0+\omega_1(1-a)$. Then what kind of
potentials can naturally give the oscillating EoS? In the
Ref.\cite{scott}, the authors built an example quintessence model,
which has the potential $V(\phi)=V_0\exp(-\lambda\phi\sqrt{8\pi
G})[1+A\sin(\nu\phi\sqrt{8\pi G})]$, where $\lambda$, $A$ and $\nu$
are all the constant numbers. They found that this model can indeed
give an oscillating EoS, if choosing appropriate parameters. In this
part, we will study the general characters of these models by
constructing potential functions from the parametrized oscillating
EoS. This method has been advised by Guo et al. in
Ref.\cite{construct}. First, we will give a simple review of this
method.

The lagrangian density of the quintessence is
 \be
 \L_{\phi}=\frac{1}{2}\dot{\phi}^2-V(\phi).
 \ee
and the pressure, energy density and EoS are
 \be
 p_{\phi}=\frac{1}{2}\dot{\phi}^2-V(\phi),~~~~~
 \rho_{\phi}=\frac{1}{2}\dot{\phi}^2+V(\phi),
 \ee
 \be
 \omega_{\phi}\equiv\frac{p_{\phi}}{\rho_{\phi}}=
 \frac{\dot{\phi}^2-2V(\phi)}{\dot{\phi}^2+2V(\phi)}
 \ee
respectively. When the energy transformation is from kinetic energy
to potential energy, the value of $\omega_{\phi}$ is damping, and on
the contrary, when the energy transformation is from potential
energy to kinetic energy, the value of $\omega_{\phi}$ is raising.
So the evolution of $\omega_{\phi}$ reflects the energy
transformation relation of the quintessence field. This suggests the
fact that it is impossible to get an oscillating EoS from the
monotonic potentials, where the quintessence fields trend to run to
the minimum of their potentials.

Consider the Flat-Robertson-Walker (FRW) universe, which is
dominated by the non-relativistic matter and a spatially homogeneous
quintessence field $\phi$. From the expression of the pressure and
energy density of the quintessence field, we have
 \be
 V(\phi)=\frac{1}{2}(1-\omega_{\phi})\rho_{\phi},
 \ee
 \be
 \frac{1}{2}\dot{\phi}^2=\frac{1}{2}(1+\omega_{\phi})\rho_{\phi}.
 \ee
These two equations relate the potential $V$ and field $\phi$ to
the only function $\rho_{\phi}$. So the main task below is to
build the function form $\rho_{\phi}(z)$ from the parametrized
EoS. This can be realized by the energy conservation equation of
the quintessence field
 \be
 \dot{\rho_{\phi}}+3H(\rho_{\phi}+p_{\phi})=0,
 \ee
where $H$ is the Hubble parameter, which yields
 \be\label{58}
 \rho_{\phi}(z)=\rho_{\phi0}\exp\left[3\int_0^z(1+\omega_{\phi})d\ln(1+z)\right]\equiv
 \rho_{\phi0}E(z),
 \ee
where $z$ is the redshift which is given by $1+z=a_0/a$ and
subscript $0$ denotes the value of a quantity at the redshift
$z=0$ (present). In term of $\omega_{\phi}(z)$, the potential can
be written as a function of the redshift $z$:
 \be
 V[\phi(z)]=\frac{1}{2}(1-\omega_{\phi})\rho_{\phi0}E(z).
 \ee
With the help of the Friedmann equation
 \be
 H^2=\frac{\kappa^2}{3}(\rho_m+\rho_{\phi}),
 \ee
where $\kappa^2=8\pi G$ and $\rho_m$ is the matter density, one
can get
 \be
 \tilde{V}[\phi]=\frac{1}{2}(1-\omega_{\phi})E(z),
 \ee
 \be
 \frac{d\tilde{\phi}}{dz}=\mp
 \sqrt{3}\frac{1}{(1+z)}
 \left[\frac{(1+\omega)E(z)}{r_0(1+z)^3+E(z)}\right]^{1/2},
 \ee
where we have defined the dimensionless quantities $\tilde{\phi}$
and $\tilde{V}$ as
 \be
 \tilde{\phi}\equiv\kappa\phi,~~~~~\tilde{V}\equiv V/\rho_{\phi0},
 \ee
and $r_0\equiv\Omega_{m0}/\Omega_{\phi0}$ is the energy density
ratio of matter to quintessence at present time. The upper (lower)
sign in Eq.(11) applies if $\dot{\phi}>0 (\dot{\phi}<0)$. These
two equations relate the quintessence potential $V(\phi)$ to the
equation of state function $\omega_{\phi}(z)$. Given an effective
equation of state function $\omega_{\phi}(z)$, the construction
Eqs.(10) and (11) will allow us to construct the quintessence
potential $V(\phi)$.

Here we consider a most general oscillating EoS as
 \be
 \omega_{\phi}=\omega_0+\omega_1\sin z,
 \ee
where $|\omega_0|+|\omega_1|\leq1$ must be satisfied for
quintessence field. We choose the cosmological parameters as
$\Omega_{\phi0}=0.7$, and $\Omega_{m0}=0.3$. For the initial
condition, we choose two different sets of parameters: case 1 with
$\omega_0=-0.7$, $\omega_1=0.2$ and $\tilde{\phi}_0=1.0$; case 2
with $\omega_0=-0.4$, $\omega_1=0.5$ and $\tilde{\phi}_0=1.0$. We
plot them in Fig.[1].

But how to fix the $``\mp"$ sign in Eq.(11)? We choose the initial
condition with $d\tilde{\phi}_0/dz<0$, assuming the variety of this
sign from $``-"$ to $``+"$ exists, then on the transformation point,
(for the continue evolution of the field $\phi$) we have
$\dot{\phi}=d\tilde{\phi}/dz=0$, which follows that
$\omega_{\phi}=-1$ at this condition. Since $\omega_{\phi}>-1$ is
always satisfied in these two models we consider, there is no
transformation of the sign in Eq.(11). So the negative sign is held
for all time. In Fig.[2], we have plotted the evolution of the
potentials of the quintessence models with redshift, and in Fig.[3],
we have plotted the constructed potentials. From these figures, one
finds that although the potential functions are oscillating, but
their amplitudes are altering with field. The field always runs from
the potential with higher amplitudes to which with lower ones.

Now let's analyze the reason of the these strange potential forms.
The evolutive equation of the quintessence field is
 \be
 \ddot{\phi}+3H\dot{\phi}+V_{,\phi}=0,
 \ee
where $V_{,\phi}$ denotes $dV/d\phi$. This equation can be rewritten
as
 \be
 \ddot{\phi}+V_{,\phi}=-3H\dot{\phi}.
 \ee
If the right-hand is absent, this is an equation which describes the
motion of field $\phi$ in the potential $V(\phi)$ in the flat
space-time. The right-hand of this equation is the effect of the
expansion of the universe. Due to show clearly its effect on the
field, we consider the simplest condition with $V(\phi)$ being a
constant, if the right-hand is absent, we get that $\ddot{\phi}=0$,
and $\dot{\phi}$ keeps constant, which is the free motion of the
field. But if adding the right-hand, we get its solution
$|\dot{\phi}|\propto e^{\int{-3Hdt}}$, the velocity of the field
rapidly decreases with time. So the effect of the cosmic expansion
is a kind of resistance to the field, and this force is directly
proportionate to the velocity of the field $\dot{\phi}$. For
overcoming this resistance force and keep the kinetic energy not
being zero, the field must roll from the region with higher
amplitude to the one with lower amplitude. This is the reason why
the potential has so strange a form as shown in Figs.[2] and [3].
Since the field is always running to the relatively smaller value of
its potential, but the potential can not be smaller than zero, it is
very difficult to build the potential with EoS which oscillates
forever if without extreme fine-tuning.

\section{Three quintessence models}

From the previous section, we find the general characters of the
potentials which can follow the oscillating EoS. According to
these characters, one can find that the potential advised in
ref.\cite{scott} indeed satisfies this condition. However in that
reference, the authors found a weak fine-tuning exists in the
model for the constraint from the BBN observation. And also this
model can obviously alter the CMB anisotropy power spectrum,
compared to the standard $\Lambda$CDM model. These are bacause the
oscillation of EoS exists at the radiation-dominant stage in that
model. Here we will build another three kinds of potential
functions, which also can generate the oscillating EoS. First we
will simplify the evolutive equations of the quintessence field.
We introduce the following dimensionless variables,
 \be
 x\equiv\frac{\kappa\dot{\phi}}{\sqrt{6}H},~~y\equiv\frac{\kappa\sqrt{V}}{\sqrt{3}H},~~
 z\equiv\frac{\kappa\sqrt{\rho_m}}{\sqrt{3}H},~~u\equiv\frac{\sqrt{6}}{\kappa\phi},
 \ee
then the evolutive equations of the matter and quintessence can be
rewritten as\cite{simply}
 \ba
 x'&=&3x(x^2+z^2/2-1)-f(y,u); \\
 y'&=&3y(x^2+z^2/2)+f(y,u)x/y;\\
 z'&=&3z(x^2+z^2/2-1/2);\\
 u'&=&-xu^2,
 \ea
where a prime denotes derivative with respect to the so-called
e-folding time $N\equiv\ln a$, and the function
$f(y,u)=\frac{\kappa V_{,\phi}}{\sqrt{6}H^2}$, which have the
different forms for different potential functions. In this
section, we mainly discuss three simple models, which have the
similar potentials as in Fig.[3]:


Model 1: $V(\phi)=V_0(\kappa\phi)^{-2}[\cos(\phi/\phi_c)+2]$ with
$\kappa\phi_c=0.1$ and
 \be
 f(y,u)=-uy^2-5\sqrt{6}y^2\sin(10\sqrt{6}/u)/[\cos(10\sqrt{6}/u)+2];
 \ee

Model 2: $V(\phi)=V_0(\kappa\phi)^{-1}[\cos(\phi/\phi_c)+2]$ with
$\kappa\phi_c=0.1$ and
 \be
 f(y,u)=-uy^2/2-5\sqrt{6}y^2\sin(10\sqrt{6}/u)/[\cos(10\sqrt{6}/u)+2];
 \ee

Model 3: $V(\phi)=V_0[(\kappa\phi)^{-1}+\cos(\phi/\phi_c)+1]$ with
$\kappa\phi_c=0.1$ and
 \be
 f(y,u)=-3y^2[u^2/6+10\sin(10\sqrt{6}/u)]/[u+\sqrt{6}+\sqrt{6}\cos(10\sqrt{6}/u)].
 \ee
These models have been shown in Fig.[4], which all are the
combinations of the invert power law function and the PNGB field.
And $V(\phi)>0$ is satisfied for all time. When $\phi/\phi_c\ll1$,
they are like the invert power law potential with $n=-1($ or
$-2)$, and they begin to oscillate when $\phi>\phi_c$. The
oscillating amplitudes decrease for all time for the first two
potentials, and for the last potential, the oscillating amplitudes
are nearly a constant at $\phi\gg\phi_c$. It is interesting that
these potentials can be looked as the invert power law potential
$3V_0(\kappa\phi)^{-1}$ $(3V_0(\kappa\phi)^{-2},
~V_0[(\kappa\phi)^{-1}+2])$ with an oscillating amendatory term at
$\phi>\phi_c$. Here we choose the initial condition (present
values) $\kappa\phi_0=0.6$, $\omega_{\phi0}=-0.9$,
$\Omega_{\phi0}=0.7$ and $\Omega_{m0}=0.3$. So at the early stage,
the potential functions of the quintessence are monotonic
function, the EoS are not oscillating at the early
(radiation-dominant) stage, which naturally overcome the
shortcoming of the model in Ref.\cite{scott}.

In Figs.[5] and [6], we plot the evolution of EoS and field $\phi$
in the region $\ln a/a_0=[0,4]$. The solid lines are the model
with the first potential, whose EoS has a relatively steady
oscillating amplitude. This is for the amplitudes in its potential
function is rapidly decreasing with $\phi$. When the field $\phi$
rolls down to its valley, it has enough kinetic energy to climb up
to its following hill and then rolls down again. At every period
of its potential, when the field is rolling down, the kinetic
energy is increasing, and the potential energy is decreasing,
which makes its EoS is raising; on the contrary, when the field is
climbing up, the kinetic energy is decreasing, and the potential
energy is increasing, which makes its EoS is damping. The minimum
values of its EoS never get $-1$, which is for the kinetic energy
of the field never get zero. This process keeps until $\ln
a/a_0\simeq1.7$ ($\kappa\phi\simeq2.2$), when the field gets the
state with $\dot{\phi}=0$ ($\omega_{\phi}=-1$), and has to return
to roll down the former valley ($\dot{\phi}<0$). These can be seen
clearly in Fig.[6]. After this state, the EoS will rapidly run to
a steady state with $\omega_{\phi}=-1$.

However all these are different for models 2 and 3, which are
described with dash and dot lines in these figures. When the fields
roll down to the valley with $\kappa\phi\simeq1$, they try to climb
up their first hills, however can not climb up to the peaks for the
large values of their potential functions. When the fields get the
state with $\dot{\phi}=0$, (the corresponding EoS are
$\omega_{\phi}=-1$) they have to return to roll down this valley
again. This process lasts until the kinetic energy become
negligible, the fields stay at the valley with $\omega_{\phi}=-1$.
The evolution of these fields can be seen clearly in Fig.[6].

In Fig.[7], we plot the evolution of $\Omega_{\phi}$ in the
universe, although the quintessence will be dominant at last in the
universe, the values of $\Omega_{\phi}$ are oscillating at the
evolution stage for all these three quintessence models, which are
determined by the evolution of $\omega_{\phi}$. When
$\omega_{\phi}>0$, the values of $\Omega_{\phi}$ will decrease, and
when $\omega_{\phi}<0$, the values of $\Omega_{\phi}$ will increase.

\section{Conclusion}

Understanding the physics of the dark energy is one of the most
mission for the modern cosmology. Until recently, the most effective
way is to detect its EoS and the running behavior  by the
observations on SNIa, CMB ,LSS and so on. There are mild evidences
to show that the EoS of the dark energy is the oscillating function,
which makes the building of the dark energy models difficult. For
the quintessence field dark energy models, it is obvious that this
EoS can't be realized from the monotonic potentials. However for the
simple oscillating potential, it is also difficult to realize.

In this paper, we have discussed the general features of the
potentials which can follow an oscillating EoS by constructing the
potentials from oscillating EoS, and found that they are
oscillating functions, however the oscillating amplitudes are
increasing (decreasing) with the field $\phi$. And also the field
must roll from the region with larger amplitude to which with
smaller amplitude if the EoS is oscillating. This kind of
potentials are not very difficult to satisfy. However since the
field must roll down to the region with smaller amplitude if the
EoS is oscillating, and also the constraint of $V(\phi)\geq0$ must
be satisfied for all time, which lead to the building of
quintessence with oscillating (forever) EoS is very difficult. In
this paper, we have studied three kinds of models:
$V(\phi)=V_0(\kappa\phi)^{-2}[\cos(\phi/\phi_c)+2]$,~
$V(\phi)=V_0(\kappa\phi)^{-1}[\cos(\phi/\phi_c)+2]$ and
$V(\phi)=V_0[(\kappa\phi)^{-1}+\cos(\phi/\phi_c)+1]$. They are all
made of the invert power law functions and the oscillating
functions, and can indeed follow the oscillating EoS, however this
oscillating behavior can only keep a finite period in all these
three models.

\begin{figure}
\centerline{\includegraphics[width=15cm]{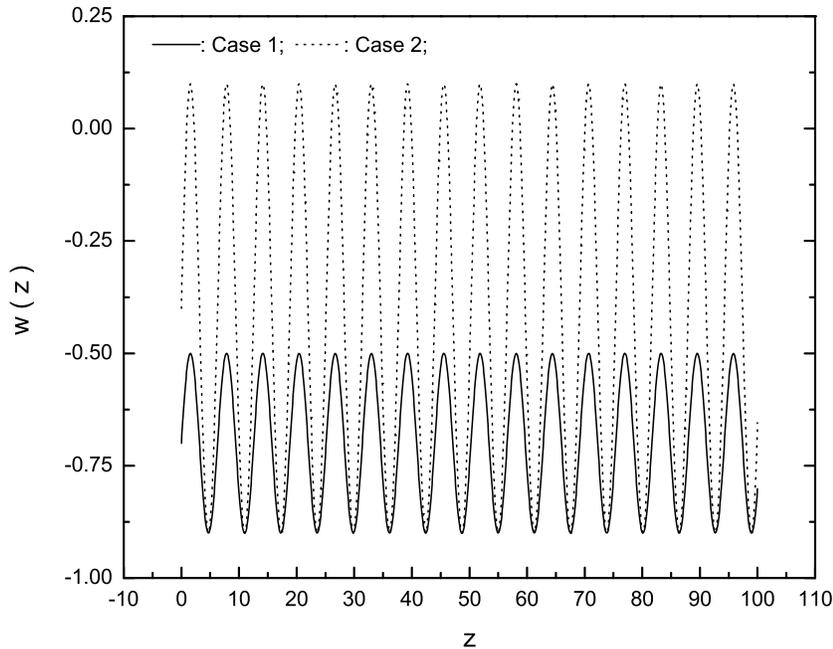}}
\caption{\small The parametrized EoS
$\omega_{\phi}(z)=\omega_0+\omega_1\sin z$.}
\end{figure}

\begin{figure}
\centerline{\includegraphics[width=15cm]{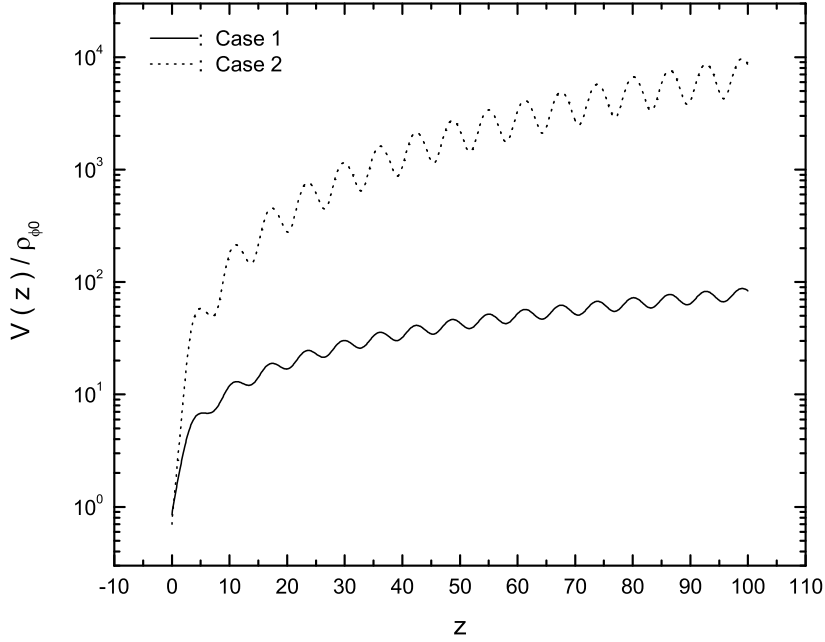}}
\caption{\small The evolution of the potentials of the
quintessence models with redshift $z$.}
\end{figure}

\begin{figure}
\centerline{\includegraphics[width=15cm]{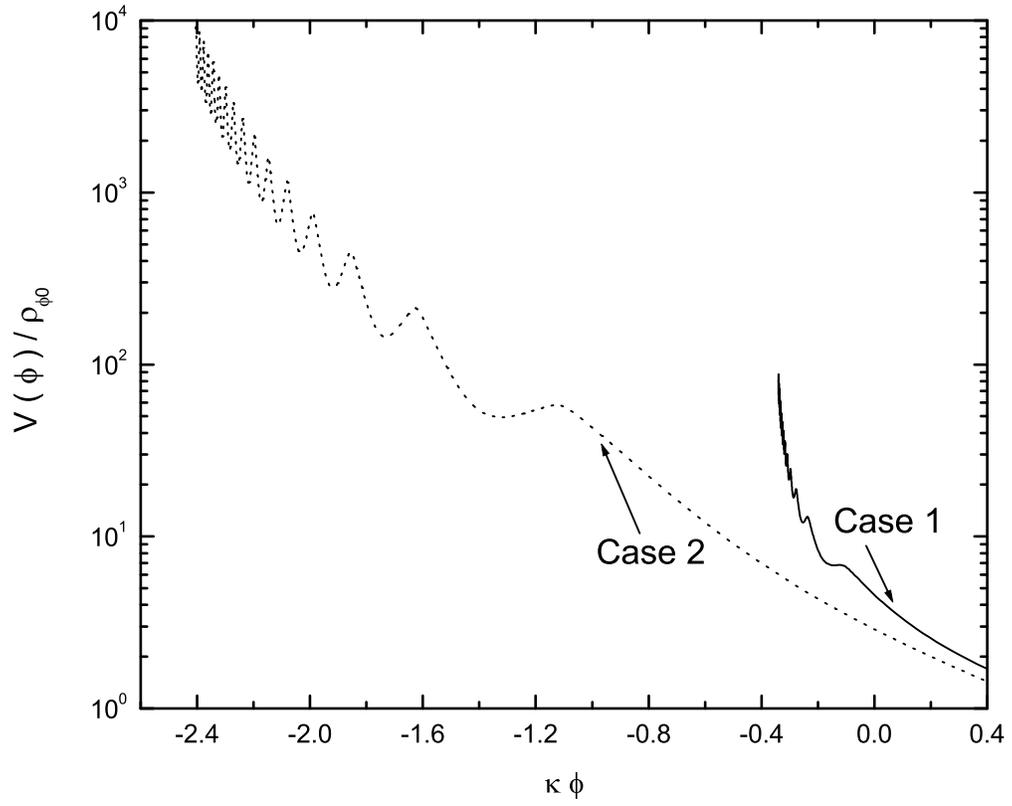}}
\caption{\small Constructed potential functions}
\end{figure}

\begin{figure}
\centerline{\includegraphics[width=15cm]{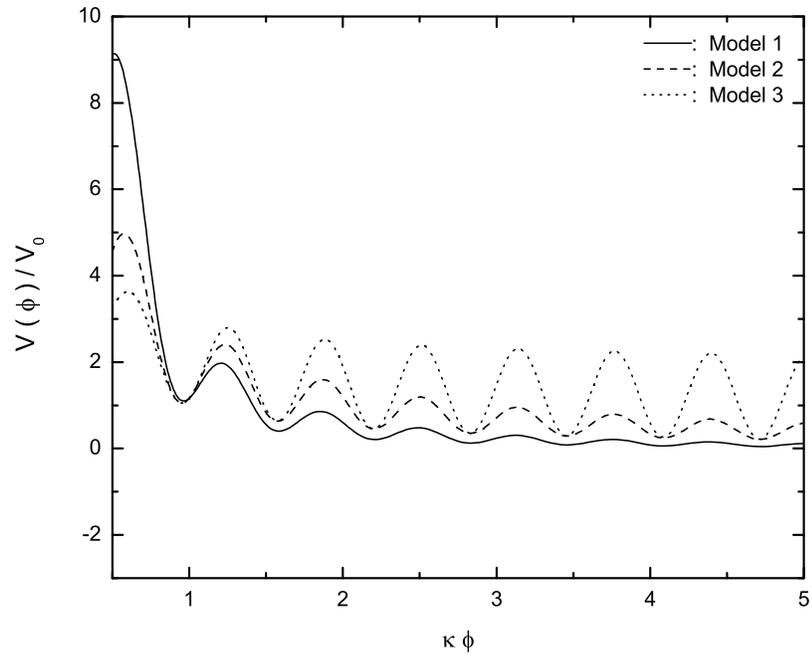}}
\caption{\small Three kinds of quintessence models. }
\end{figure}

\begin{figure}
\centerline{\includegraphics[width=15cm]{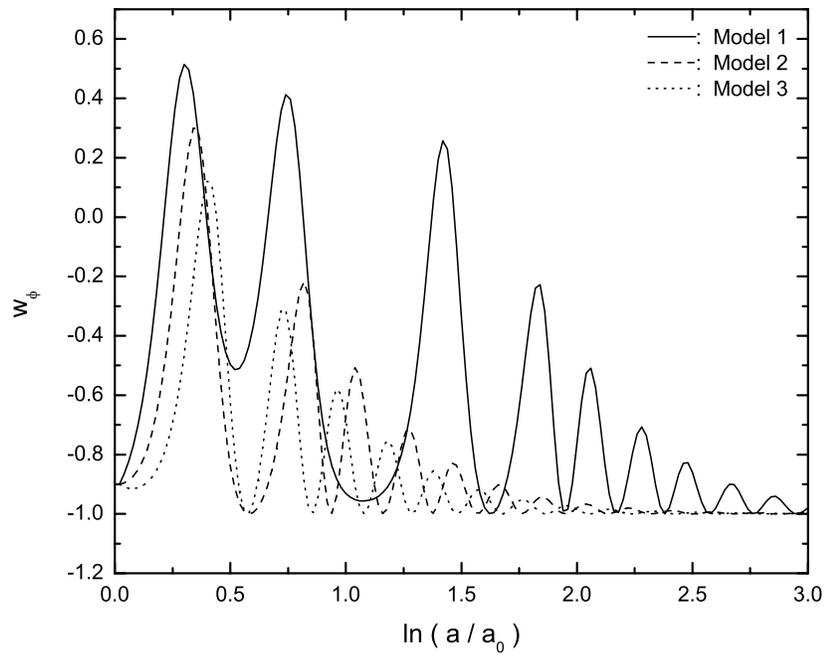}}
\caption{\small The evolution of the EoS of the quintessence
models.}
\end{figure}

\begin{figure}
\centerline{\includegraphics[width=15cm]{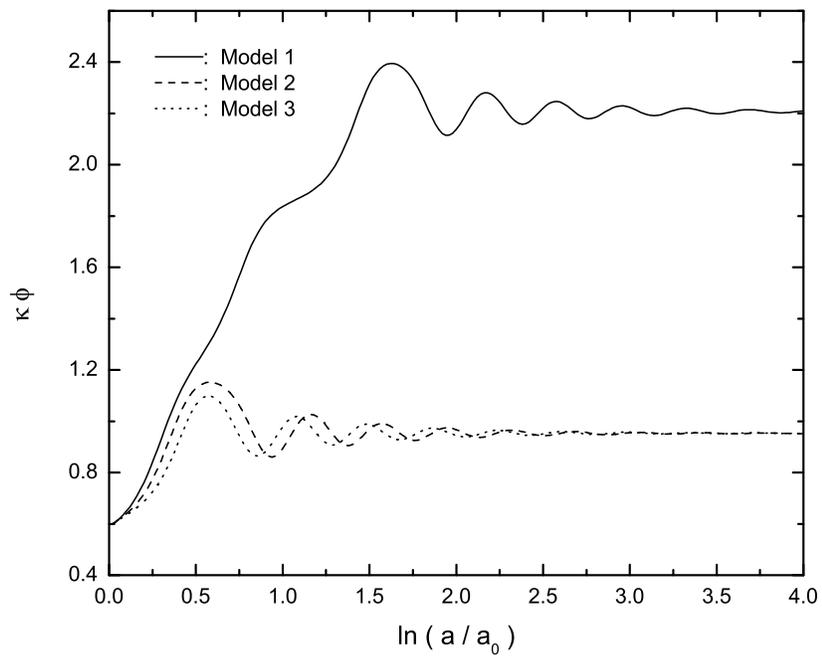}}
\caption{\small The evolution of field $\phi$ of the quintessence
models.}
\end{figure}

\begin{figure}
\centerline{\includegraphics[width=15cm]{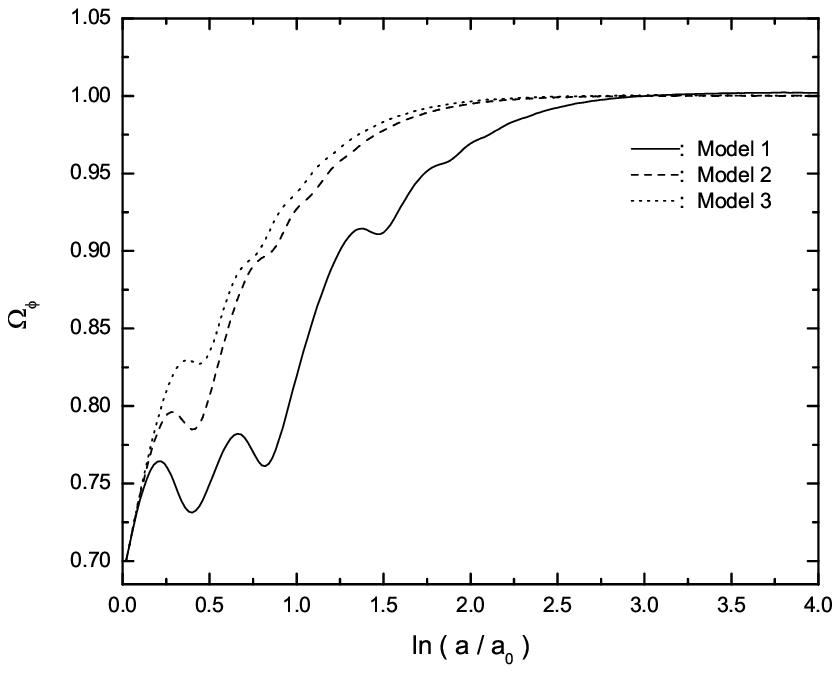}}
\caption{\small The evolution of the energy density
$\Omega_{\phi}$ of the quintessence models.}
\end{figure}



\baselineskip=12truept

\begin{thebibliography}{99}

\bibitem{sn}
Riess A.G. {\it et al.}, Astron.J. {\bf 116} (1998) 1009;
Perlmutter S. {\it et al.}, Astrophys.J. {\bf 517} (1999) 565;
Tonry J.L. {\it et al.}, Astrophys.J. {\bf 594} (2003) 1; Knop
R.A. {\it et al.}, Astrophys.J. {\bf 598} (2003) 102;


\bibitem{map}
Bennett C.L. \emph{et al.}, Astrophys.J.Suppl. {\bf 148} (2003) 1;
Spergel D.N. \emph{et al.}, Astrophys.J.Suppl. {\bf 148} (2003)
175; Peiris H.V. \emph{et al.}, Astrophys.J.Suppl. {\bf 148}
(2003) 213; Spergel D.N. \emph{et al.}, astro-ph/0603449;


\bibitem{sdss}
Tegmark M. \emph{et al.}, Astrophys.J. {\bf 606} (2004) 702,
Phys.Rev.D {\bf 69} (2004) 103501; Pope A.C. \emph{et al.},
Astrophys.J. {\bf 607} (2004) 655; Percival W.J. \emph{et al.},
MNRAS {\bf 327} (2001) 1297;


\bibitem{phantom}
Caldwell R.R., Phys.Lett.B {\bf 545} (2002) 23; Carroll S.M.,
Hoffman M. and Trodden M., Phys.Rev.D {\bf 68} (2003) 023509;
R.R.caldwell, M.Kamionkowski and N.N.Weinberg, Phys.Rev.Lett. {\bf
91} (2003) 071301;

\bibitem{kessence}
Armendariz C., Damour T. and Mukhanov V., Phys.Lett.B {\bf 458}
(1999) 209; Chiba T., Okabe T. and Yamaguchi M., Phys.Rev.D {\bf
62} (2000) 023511; Armendariz C., Mukhanov V. and Steinhardt P.J.,
Phys.Rev.D {\bf 63} (2001) 103510;


\bibitem{quintom}
Feng B., Wang X.L. and Zhang X.M., Phys.Lett.B {\bf 607} (2005) 35;
Guo Z.K., Piao Y.S., Zhang X.M. and Zhang Y.Z., Phys.Lett.B {\bf
608} (2005) 177; Hao W. and Cai R.G., Phys.Rev.D {\bf 72} (2005)
123507; Hu W., Phys.Rev.D {\bf 71} (2005) 047301;


\bibitem{yangmills}
Zhang Y., Chin.Phys.Lett. {\bf 20} (2003) 1899; Zhao W. and Zhang
Y., Class.Quant.Grav. {\bf 23} (2006) 3405; Phys.Lett.B {\bf 640}
(2006) 69; astro-ph/0508010; Zhang Y., Xia T.Y. and Zhao W.,
gr-qc/0609115;


\bibitem{quint}
Wetterich C., Nucl.Phys.B {\bf 302} (1988) 668 ; Astron.Astrophys.
{\bf 301} (1995) 321; Ratra B. and Peebles P.J., Phys.Rev.D {\bf
37} (1988) 3406; Caldwell R.R., Dave R. and Steinhardt P.J.,
Phys.Rev.Lett. {\bf 80} (1998) 1582; Zhai X.H. and Zhao Y.B.
Chin.Phys. {\bf 15} (2006) 1009;

\bibitem{simply}
Copeland E.J., Liddle A.R. and Wands D., Phys.Rev.D {\bf 57}
(1998) 4686; Amendola L., Phys.Rev.D {\bf 60} (1999) 043501;
Amendola L., Phys.Rev.D {\bf 62} (2000) 043511



\bibitem{track_quint}
Zlatev I., Wang L. and Steinhardt P.J. Phys.Rev.Lett. {\bf 82}
(1999) 896; Steinhardt P.J., Wang L. and Zlatev I., Phys.Rev.D
{\bf 59} (1999) 123504;


\bibitem{scott}
Dodelson S., Kaplinghat M. and Stewart E., Phys.Rev.Lett. {\bf 85}
(2000) 5276;


\bibitem{osci}
Feng B., Li M.Z. and Zhang X.M., Phys.Lett.B {\bf 634} (2006) 101;
Xia J.Q., Feng B. and Zhang X.M., Mod.Phys.Lett.A {\bf 20} (2005)
2409; Barenboim G., Mena O. and Quigg C., Phys.Rev.D {\bf 71}
(2005) 063533; Barenboim G. and Lykken J., Phys.Lett.B {\bf 633}
(2006) 453; Linder E.V., Astropart.Phys. {\bf 25} (2006) 167; Zhao
W. and Zhang Y., Phys.Rev.D {\bf 73} (2006) 123509;


\bibitem{s-t}
Nojiri S. and Odintsov S.D., Phys.Lett.B {\bf 637} (2006) 139;


\bibitem{observation}
Huterer D. and Cooray A., Phys.Rev.D {\bf 71} (2005) 023506; Lazkoz
R., Nesseris S. and Perivolaropoulos L., JCAP {\bf 0511} (2005) 010;
Xia J.Q., Zhao G.B., Li H., Feng B. and Zhang X.M., Phys.Rev.D {\bf
74} (2006) 083521;

\bibitem{pngb}
Freese K., Frieman J.A. and Olinto A.V., Phys.Rev.Lett. {\bf 65}
(1990) 3233; Asams F.C., Bond J.R., Freese K., Frieman J.A. and
Olinto A.V., Phys.Rev.D {\bf 47} (1993) 426; Frieman J., Hill C.,
Stebbins A. and Waga I., Phys.Rev.Lett. {\bf 75} (1995) 2077;
Copeland E.J., Sami M. and Tsujikawa S., hep-th/0603057;

\bibitem{construct}
Guo Z.K., Ohta N. and Zhang Y.Z., Phys.Rev.D {\bf 72} (2005) 023504;
Li Hui, Guo Z.K. and Zhang Y.Z., Mod.Phys.Lett.A {\bf 21} (2006)
1683; Cao H.M., Xin X.B, Wang L. and Zhao W., Journal of Science and
Technology of China accepted;





\end{thebibliography}
\end{document}